\documentclass[journal=jacsat,manuscript=article]{achemso}

\usepackage{amssymb}
\usepackage{graphicx}
\usepackage{dcolumn}
\usepackage[version=3]{mhchem}
\usepackage{textcomp, gensymb}

\author{Jonathan G. Raybin}
\affiliation{Department of Chemistry, University of California, Berkeley, CA 94720, United States.}

\author{Rebecca B. Wai}

\author{Naomi S. Ginsberg}
\email{nsginsberg@berkeley.edu}
\affiliation{Department of Chemistry, University of California, Berkeley, CA 94720, United States.}
\alsoaffiliation{Department of Physics, University of California, Berkeley, CA 94720, United States.}
\alsoaffiliation{Molecular Biophysics and Integrated Bioimaging Division, Lawrence Berkeley National Laboratory, Berkeley, CA 94720, United States.}
\alsoaffiliation{Materials Sciences Division, Lawrence Berkeley National Laboratory, Berkeley, CA 94720, United States.}
\alsoaffiliation{Kavli Energy NanoScience Institute, Berkeley, CA 94720, United States.}
\alsoaffiliation{STROBE, NSF Science \& Technology Center, Berkeley, California 94720, United States.}

\title{Non-Additive Interactions Unlock Small-Particle Mobility in Binary Colloidal Monolayers}

\begin{document}

\begin{tocentry}

\includegraphics[width=5.8cm]{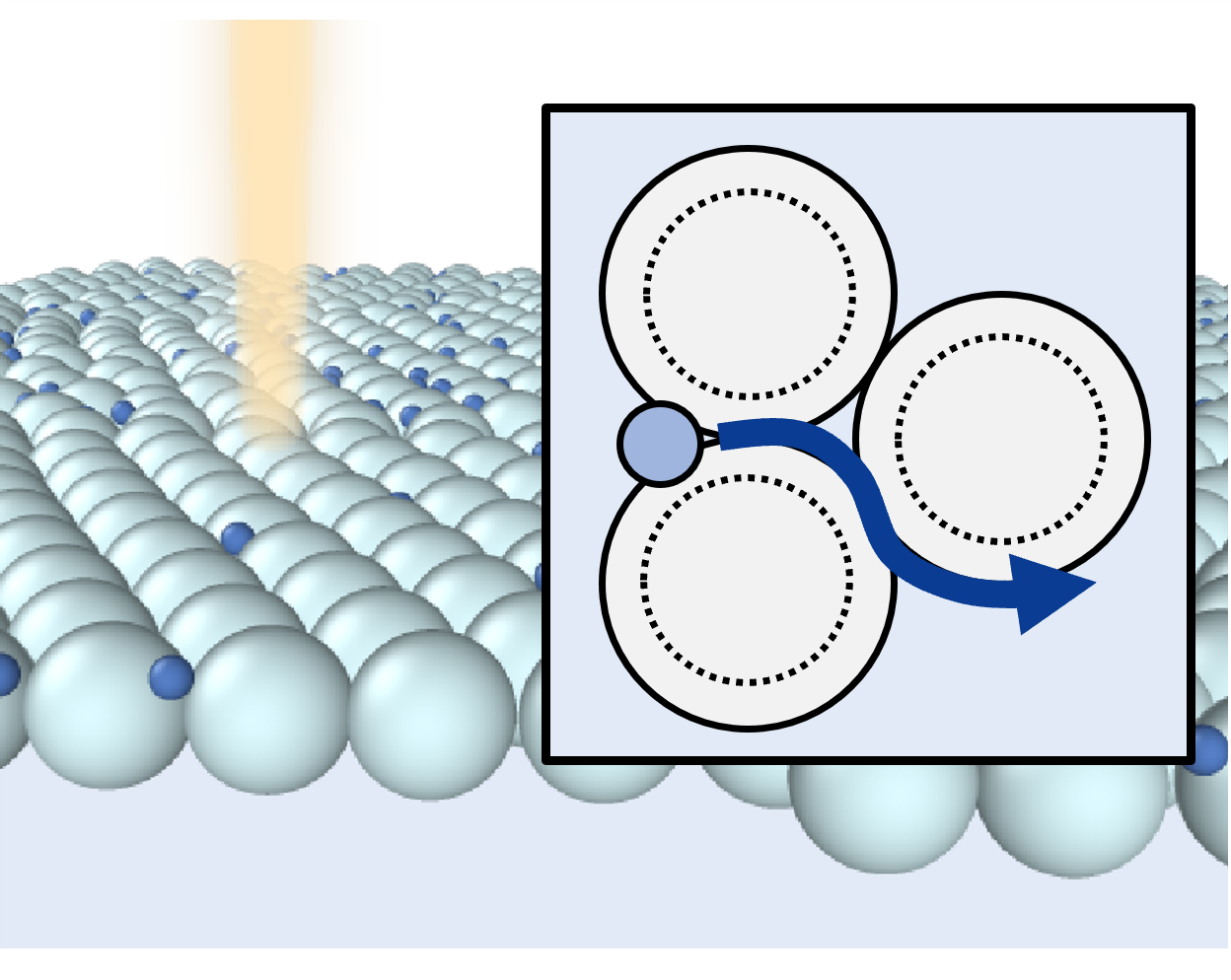}

\end{tocentry}

\begin{abstract}

We examine the organization and dynamics of binary colloidal monolayers composed of micron-scale silica particles interspersed with smaller-diameter silica particles that serve as minority component impurities. These binary monolayers are prepared at the surface of ionic liquid droplets over a range of size ratios ($\sigma=0.16-0.66$) and are studied with low-dose minimally perturbative scanning electron microscopy (SEM). The high resolution of SEM imaging provides direct tracking of all particle coordinates over time, enabling a complete description of the microscopic state. In these bidisperse size mixtures, particle interactions are non-additive because interfacial pinning to the droplet surface causes the equators of differently sized particles to lie in separate planes. By varying the size ratio we control the extent of non-additivity in order to achieve phase behavior inaccessible to additive 2D systems. Across the range of size ratios we tune the system from a mobile small-particle phase ($\sigma<0.24$), to an interstitial solid ($0.24<\sigma<0.33$), to a disordered glass ($\sigma>0.33$). These distinct phase regimes are classified through measurements of hexagonal ordering of the large-particle host lattice and the lattice’s capacity for small-particle transport. Altogether, we explain these structural and dynamic trends by considering the combined influence of interparticle interactions and the colloidal packing geometry. Our measurements are reproduced in molecular dynamics simulations of 2D non-additive disks, suggesting an efficient method for describing confined systems with reduced dimensionality representations.

\end{abstract}

\subsubsection{Keywords:}
colloids, soft condensed matter, electron microscopy, interfacial assembly, ionic liquid, nanoparticles, non-additivity

\maketitle
\section*{Introduction}

Colloidal nanoparticles serve as versatile building blocks for the self-assembly of nanostructured materials, due to the tunability of their material, size, shape, and surface chemistry.\cite{boles_self-assembly_2016, manoharan_colloidal_2015} Combining multiple colloidal components introduces additional length and energy scales that further expand the scope of possible structures.\cite{thorneywork_structure_2018, van_der_meer_diffusion_2017, cho_self-organization_2005} Even in the relatively simple case of two-dimensional (2D) binary mixtures of differently-sized spheres, particle assemblies exhibit a diverse array of morphologies, including crystalline, \cite{dong_two-dimensional_2011, lotito_approaches_2017} quasicrystalline, \cite{talapin_quasicrystalline_2009, ye_quasicrystalline_2017, fayen_self-assembly_2023} and amorphous phases.\cite{gao_bidisperse_2020} In general, the assembly process is governed by a range of competing kinetic and thermodynamic effects, dependent on a hierarchy of particle-particle and particle-environment interactions. Despite this overall complexity, much of the observed structural diversity can be generated from simplified model systems of hard disks interacting only through volume exclusion.\cite{filion_prediction_2009,fayen_infinite-pressure_2020}

In bidisperse monolayers composed of large and small particles, with respective radii $r_L$ and $r_S$, in addition to the total particle density $\phi_T$ the system state also depends on the binary number fraction $\chi_S=1-\chi_L$ and size ratio $\sigma=r_S/r_L$. Additionally, in many binary mixtures the pairwise interaction length scale $2 r_{LS}$ does not necessarily correspond to the additive sum of the component radii, $r_L + r_S$. To account for this difference, binary hard-disk models may be generalized by incorporating a non-additivity term $\Delta$ such that $r_{LS}=\dfrac{1}{2}(r_L+r_S)(1+\Delta)$.\cite{dijkstra_phase_1998} Positive non-additivity introduces an effective interspecies repulsion and can lead to phase separation, while negative non-additivity generates an effective attraction that promotes mixing. In experimental systems, non-additivity frequently arises from soft interactions due to particle charge, surface chemistry, solvation, or ligand intercalation.\cite{silvera_batista_nonadditivity_2015} For example, the widely-applied Asakura-Oosawa model for depletion interactions in polymer-colloid mixtures assumes no interaction between the polymer chain depletants and represents a limiting case of positive non-additivity.\cite{binder_perspective_2014} In addition to hard-core models, analogous non-additivity relations have been developed for soft potentials, including electrostatic interactions between differently-sized charged colloidal particles. In each of these cases, an accurate understanding of the effects of non-additivity is critical, as non-additive mixtures access distinct structural phases with unique material properties.\cite{widmer-cooper_structural_2011, fayen_infinite-pressure_2020} Non-additive interactions have also recently been proposed as a design mechanism for the programmed self-assembly of monolayers with controlled open lattice architectures.\cite{salgado-blanco_non-additive_2015}

Interfacially confined monolayers serve as a natural platform for studying non-additivity in 2D systems. Whether sedimented at a solid interface or adsorbed to a fluid interface, the equators of spheres of different sizes lie at different levels from the surface. The resulting height offset leads to an effective shortening of the minimum approach distance between large and small spheres when projected onto the plane in which they make contact. Meanwhile, particles of the same size lie in a common plane, and their contact distances remain unchanged. Consequently, under the confined geometry, particle interactions may be described following a 2D non-additive representation. Following these arguments, several recent studies have simulated hard-disk mixtures with negative non-additivity to analyze the phase behavior of confined nanospheres.\cite{salgado-blanco_non-additive_2015, fayen_infinite-pressure_2020, fayen_self-assembly_2023} To our knowledge, however, the effect of non-additivity on binary assembly at interfaces has not been examined experimentally. Although the assembly and dynamics of binary colloidal monolayers have been extensively studied,\cite{baumgartl_experimental_2007, ebert_partial_2009, mazoyer_dynamics_2009, bonales_phase_2012, thorneywork_communication_2014, thorneywork_self-diffusion_2017, lavergne_equilibrium_2016, gao_bidisperse_2020, zhou_discovery_2022} these investigations have considered only single size ratios or have focused on size ratio regimes where the effects of non-additivity may be neglected.

Here, we systematically measure the structural and dynamic properties of bidisperse mixtures of interfacially-confined silica nanospheres over a range of size ratios. Despite the inherently 3D geometry, we demonstrate that the monolayer may be represented as a 2D system of non-additive particles. Over this series, non-additivity, which is enhanced with increasing size asymmetry, plays an essential role in determining monolayer properties. For example, at low size ratios we observe a mobile small-particle phase exhibiting delocalized transport properties that would be inaccessible to additive 2D systems. We also find that non-additivity reduces lattice strain, leading to improved structural ordering of the large particles over a broader range of size ratios. Although we obtain evidence for additional interparticle interactions beyond area exclusion, we find that a simple non-additive hard-disk model is sufficient for explaining the observed phase behavior. Altogether, these observations can be understood from 3D geometric packing arguments resulting from interfacial confinement. 

\section*{Results}

\begin{figure*}
\includegraphics[width=9cm]{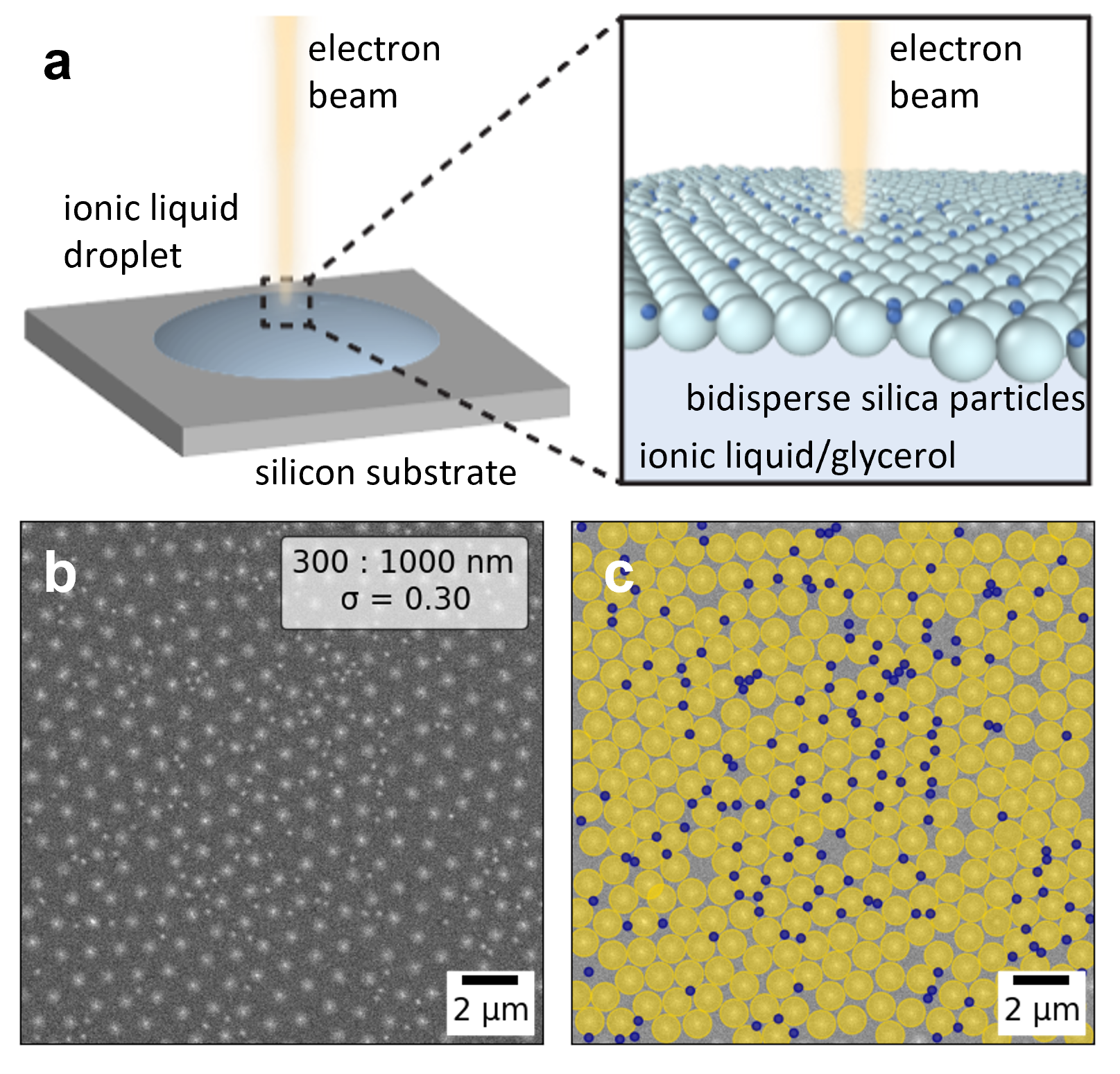}
\caption{\label{fig1} SEM monolayer imaging. (a) Experimental schematic showing assembly of  binary nanospheres at the interface of an ionic liquid droplet with the electron beam scanning over the monolayer. (b) Representative SEM image of a binary monolayer and (c) the same image labeled with space-filling circles showing the full particle sizes projected onto a 2D plane, as measured through automated particle tracking. 
}
\label{F1}
\end{figure*}

We use scanning electron microscopy (SEM) to image colloidal monolayers comprising bidisperse mixtures of silica nanospheres with varying size ratio ($\sigma=0.16-0.67$). Figure~\ref{F1}a illustrates the imaging geometry, with colloidal monolayers organized at a droplet surface. For these experiments, we suspend the particles on ionic liquid droplets, as the low vapor pressure allows direct compatibility with the SEM vacuum environment.\cite{bischak_charging-driven_2020, kim_assessing_2019, kim_visualizing_2016, gao_bidisperse_2020} Prior to imaging, droplets are allowed to equilibrate for at least 12 hours under an inert environment to facilitate the formation of dense monolayer assemblies. Following equilibration, particles collect around the droplet center in a single, large-area ($\sim 1 \rm mm^2$) patch with consistent density, surrounded by dilute particles at the periphery of the droplet. Increasing the particle concentration increases the size of this patch, and we do not find evidence for the formation of multilayers or colloidal aggregation below the monolayer surface. Within the high-density patch, particles saturate at the interface with total surface coverage fractions of $\phi_T=0.76\pm0.04$, calculated as $\phi_T=\phi_L+\phi_S-\phi_{LS}$, where $\phi_i$ is the surface coverage of each component $i=L$ or $S$, and $\phi_{LS}$ represents the large-small area overlap. Although the local composition varies across the monolayer surface, all presented imaging regions were selected with small-particle number fractions $\chi_S=0.29\pm0.04$, except where otherwise noted. These conditions correspond to dense, large-particle majority lattices where the small particles act as minority component impurities. Following their initial assembly, the high density of particle contacts hinders further rearrangements such that the resulting monolayers remain kinetically trapped in configurations that are locally stable but outside of global equilibrium. Still, average monolayer properties are consistent across regions with the same local composition and are reproducible over multiple droplet samples.

SEM imaging ensures sufficient spatial and temporal resolution to directly locate all 2D particle coordinates over time. For each size ratio, we measure time-resolved movies of particle dynamics in the monolayer with 2.1 s time resolution. We also separately acquire multiple images of the initial particle configuration over larger fields of view to assess the monolayer structure. In each of these experiments, we use a very low 15 pA beam current for minimally perturbative imaging. In this regime, particle charging from the cumulative electron dose preserves monolayer stability for roughly 2 min, which is more than sufficient for acquiring dynamic information at each pristine sample region of interest. At this low beam dosage, we observe no changes in contrast or imaging artifacts due to particle charging over the course of imaging. Further discussion of the effects of perturbation is included in the Supporting Information (Figure S1). A representative secondary electron image of a 300 nm:1000 nm binary mixture is shown in Figure~\ref{F1}b. These images were acquired with an accelerating voltage of 3 kV, such that all particles are clearly resolved and that large and small particles can be distinguished by automated particle tracking. At their equilibrium contact angle, surface-bound particles are almost entirely submerged below the liquid interface, and secondary electrons scatter from only their upper cap region.\cite{kim_visualizing_2016, bischak_charging-driven_2020} Although the particles appear to be spatially separated, labeling the image with space-filling radii for the large (yellow) and small (blue) particles in Figure~\ref{F1}c reveals direct contacts between particles. Imaging at higher accelerating voltages probes greater depths and further confirms that particles are in close contact below the ionic liquid surface, as seen in Figure S2.


\begin{figure*}
\includegraphics[width=12cm]{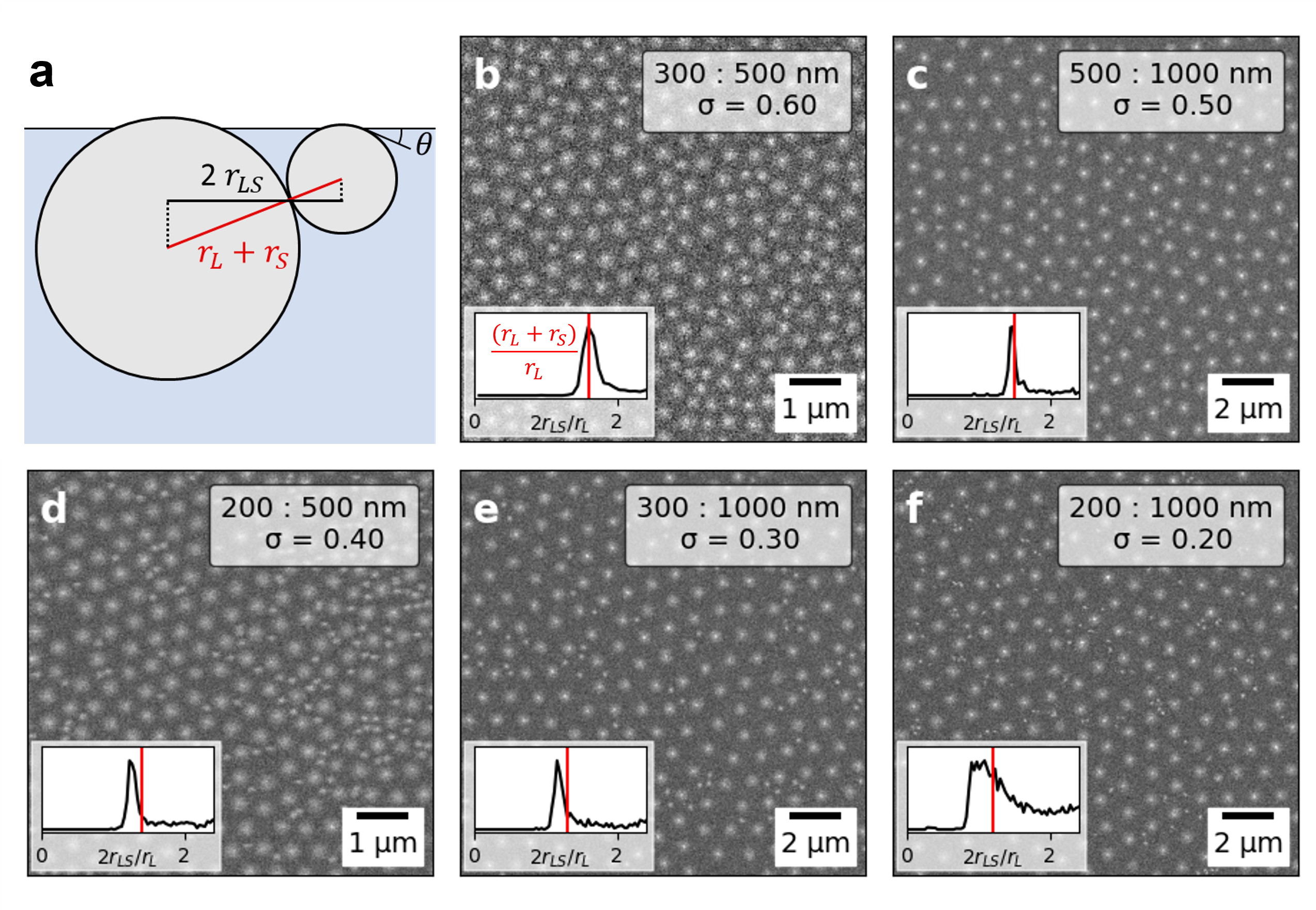}
\caption{\label{fig2} Bidisperse monolayer images over varying size ratio. (a) Schematic of the non-additive interfacial geometry showing large and small particles attached to the interface with the same contact angle $\theta$. The full center-to-center distance $r_L + r_S$ (red) is greater than the 2D projected center-to-center distance $2 r_{LS}$ (black). (b-f) Representative monolayer images with consistent $\chi_S = 0.29 \pm 0.05$ over a series of size ratios (b) 0.60, (c) 0.50, (d) 0.40, (e) 0.30, and (f) 0.20. Insets for each image plot the measured distributions of pairwise large-small separations $2 r_{LS}$ (black curve) compared to the the additive particle radii $r_L$ + $r_S$ (red vertical line), measured in units of the large-particle radius.
}
\label{F2}
\end{figure*}

\textbf{\emph{Control over Non-Additivity.}} In bidisperse monolayers, large and small particles, with respective radii $r_L$ and $r_S$, are pinned with identical contact angles to the droplet surface, displacing their equatorial planes relative to one another. As shown in Figure~\ref{F2}a, this configuration causes the projected top-down contact distances of large and small particles $2 r_{LS}$ to be shorter than the additive sum of their radii $r_L+r_S$. Representative SEM images over a series of size ratios are shown in Figure~\ref{F2}b-f, with absolute particle size ranging from 200 nm to 1 $\mu$ m.  Note that the images shown have been extracted from larger fields of view and scaled such that the large particles have similar apparent sizes. Across the full range of size ratios, monolayers form densely packed assemblies with direct contacts between most neighboring particles. Inset plots show the distribution of measured separation distances for each large-small neighbor pair as compared to the additive sum. Particle separations show broad, asymmetric distributions as not all large-small pairs reflect direct contacts, but particles cannot approach closer than the contact distance; this broadening is particularly evident at the smallest size ratios where there is more free space in the interstices between large particles. In general, the peak value of $2 r_{LS}$ is smaller than $r_L+r_S$, and the relative difference between these two quantities increases with increasing size asymmetry as the system becomes more non-additive.

\begin{figure*}
\includegraphics[width=9cm]{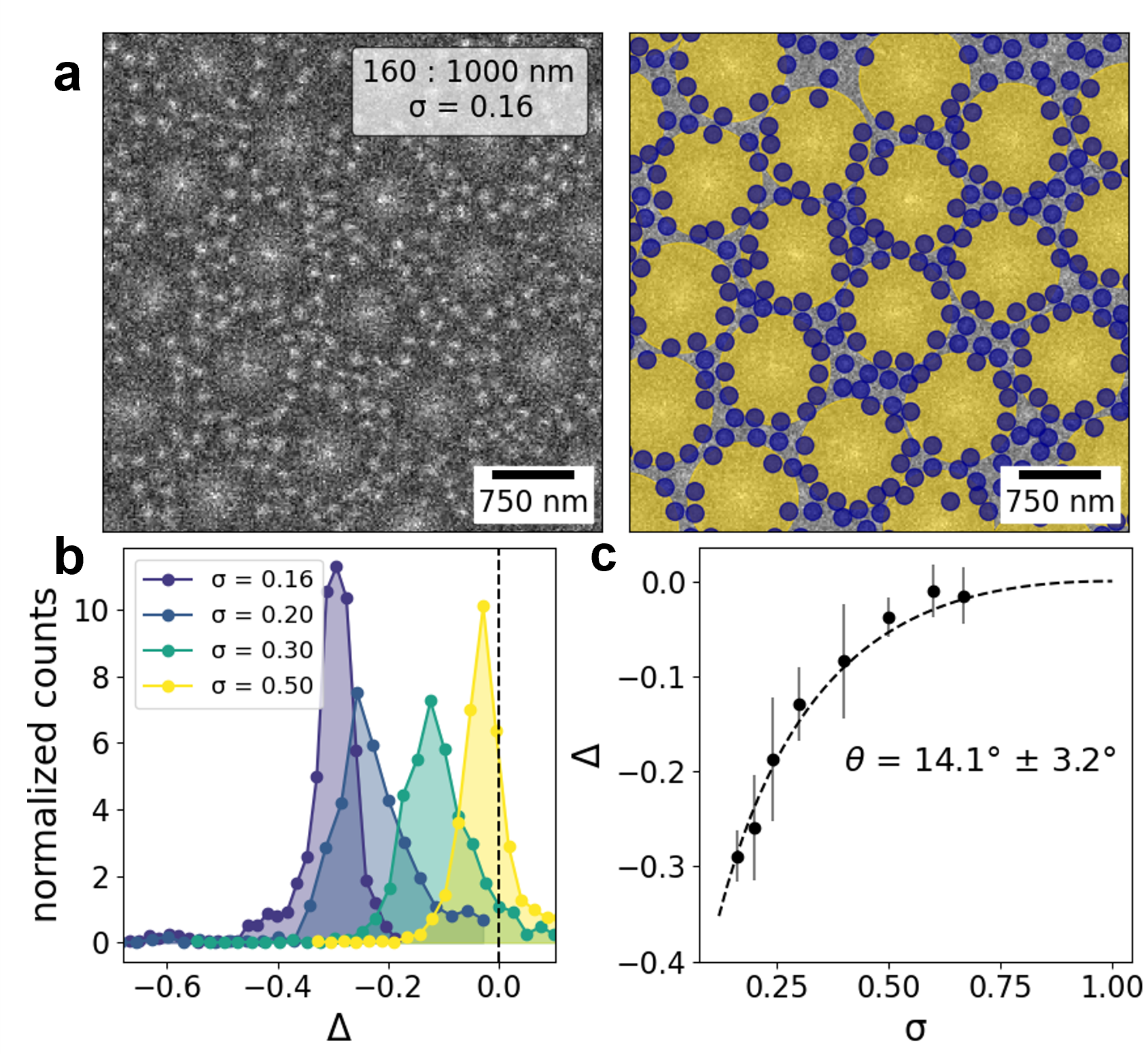}
\caption{\label{fig3} Non-additivity quantification. (a) SEM image showing particle overlap in a $\sigma=0.16$ monolayer with $\chi_S = 0.93$ and the same image labeled with space-filling circles. (b) Histograms showing the shift of measured non-additivity distributions with increasing size ratio. (c) Plot of $\Delta$ as a function of size ratio. The contact angle is measured from fitting to Equation 2 as indicated by the dashed black curve.
}
\label{F3}
\end{figure*}

This trend is reflected in observations of overlapping large and small particles in the projected 2D images. An example is shown for a $\sigma=0.16$ monolayer with higher small-particle number fraction $\chi_S$ in Figures~\ref{F3}a-b, where small particles fill the interstitial space between large particles. From measurements of $r_{LS}$ the non-additivity parameter may be quantified as: 
\begin{equation}
\Delta=\frac{2 r_{LS}}{r_L+r_S}-1. \label{EQ1}						
\end{equation}
In the case of overlapping particles the non-additivity $\Delta$ is negative. From direct measurements of $r_{LS}$ across multiple imaging regions, we determine non-additivity distributions for monolayers of each size ratio, as shown in Figure~\ref{F3}c. For these measurements, images with local compositions of $\chi_S>0.50$ were used for systems with $\sigma \leq 0.3$ to ensure adequate small-large contact statistics. The peak positions of the resulting distribution are most negative at low $\sigma$ and approach 0 with decreasing size asymmetry. Across the range of measured size ratios, non-additivity may be tuned from $\Delta=-0.29 \pm 0.03$ at $\sigma=0.16$ to $\Delta=-0.02 \pm 0.03$ at $\sigma=0.67$.

The peak positions of $\Delta$ are plotted as a function of $\sigma$ in Figure~\ref{F3}d, with uncertainties corresponding to the full widths at half maximum. Under the assumption that large and small silica particles attach to the interface with the same contact angle $\theta$, the non-additivity may be determined geometrically as a function of the size ratio and contact angle: 
\begin{equation}
\Delta=\sqrt{1-\frac{(\sigma-1)^2}{(\sigma+1)^2}\cos^2 \theta}-1. \label{EQ2}	
\end{equation}
In this expression, $\Delta$ vanishes at the additive limit of $\theta=90 \degree$, where particle centers collect at the interfacial plane. At $\theta=0 \degree$, where particles lie tangent to the interface, $\Delta$ is most negative and reduces to $\Delta=\frac{2 \sqrt{2}}{1+\sigma}-1$. This limit also corresponds to the case of particle sedimentation at a rigid interface.\cite{fayen_infinite-pressure_2020} Our measured values of $\Delta$ agree closely with Equation \ref{EQ2}, indicated by the black curve in Figure~\ref{F3}d, suggesting that the interfacial geometry shown in Figure~\ref{F2}a provides a descriptive model of our system. Because $\theta$ serves as the only adjustable parameter, this fit enables an independent measure of the particle contact angle from plane-view SEM images. We estimate $\theta=14.1 \pm 3.1 \degree$, consistent with previous measurements of silica nanoparticles in ionic liquid in the range $\theta=12\degree$ - $15\degree$.\cite{kim_assessing_2019} Consequently, this non-additive representation serves as a practical framework for reconstructing the organization of the inherently 3D system using 2D imaging data.

\begin{figure*}
\includegraphics[width=13 cm]{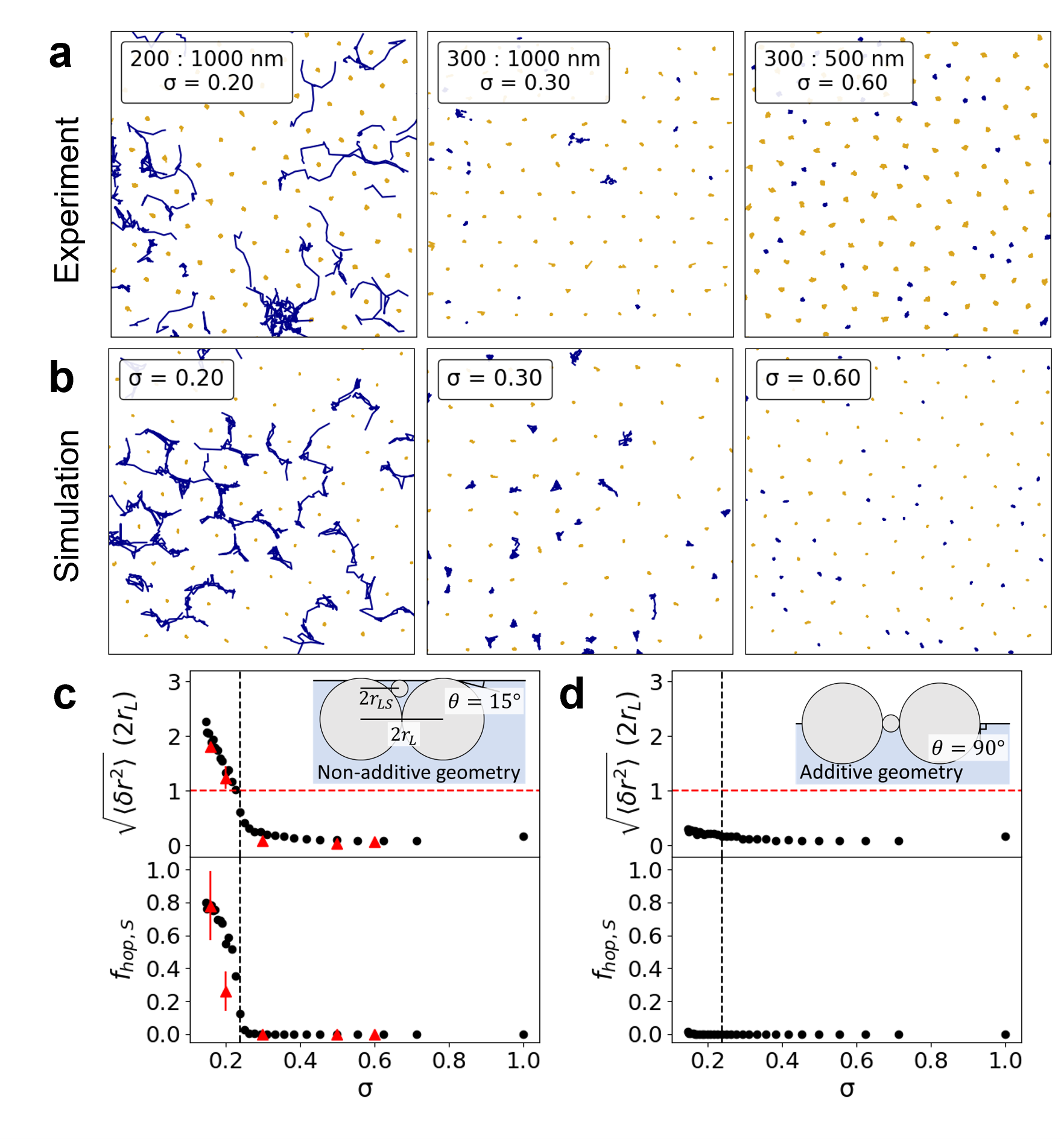}
\caption{\label{fig4} Monolayer dynamics. (a) Experimentally measured particle trajectories for a series of size ratios $\sigma=0.20$, $0.30$, and $0.60$ over imaging times of 88 s. Small-particle trajectories are shown in blue, and large-particle trajectories are in yellow. (b) Trajectories measured from simulations of non-additive WCA particles for the same size ratios. (c) Plots measuring particle dynamics as a function of size ratio based on experiment and non-additive simulations. The inset depicts the effective non-additive geometry used for the simulations, corresponding to a contact angle of $15\degree$. The upper panel shows the RMS displacement of small particles for experiment (red triangles) and simulation (black circles) after a delay time $\tau$. The lower panel measures the fraction of small particles with a displacement greater than one lattice spacing. An abrupt decrease in small-particle mobility is observed as $\sigma$ increases past 0.24, indicated by the dashed vertical line. (d) Similar plots measuring the dynamics of simulated additive disks. The inset shows the effective geometry of the additive system, corresponding to a contact angle of $90\degree$. No small-particle mobility crossover is observed in additive simulations.
}
\label{F4}
\end{figure*}

\textbf{\emph{Size Ratio-Dependent Small-Particle Mobility.}} Having established the relationship between non-additivity $\Delta$ and the binary size ratio $\sigma$, we investigate the influence of non-additivity on particle mobility. Figure~\ref{F4}a shows experimental trajectories of large (yellow) and small (blue) particles obtained from SEM movie data over a series of three representative size ratios. At each size ratio, large particles exhibit minimal displacements over experimental time scales due to the high density of large-particle contacts. The large-particle network therefore serves as a stable reference environment for tracking the relative dynamics of small particle impurities. At low size ratios, as shown for $\sigma=0.20$, small particles are observed to percolate through the interstices of the large-particle lattice. At the low small-particle density $\chi_S = 0.29 \pm 0.05$ studied here, most interstitial sites are unoccupied such that small particles can travel between sites without interacting. At $\sigma=0.30$, small-particle mobility is suppressed and trajectories become trapped within individual large-particle hollow-site cages. Near vacancies and defect sites in the lattice, small-particle trajectories at this size ratio explore the larger accessible free space but remain locally confined by the cage structure. With further increases in size ratio, as seen for $\sigma=0.60$, small-particle mobility becomes fully restricted and yields compact, localized trajectories. In this regime, large and small particles together form a common packing network that limits the mobility of both species.

As a test of this point, we find that the observed trajectories may be recapitulated through molecular dynamics simulations of non-additive particles with steep short-range repulsion described by the Weeks-Chandler-Andersen (WCA) potential.\cite{weeks_role_1971} At each size ratio, simulated trajectories, as shown in Figure~\ref{F4}b, closely resemble those of the corresponding experimental monolayer. In these simulations, the non-additivity $\Delta$ was determined from the binary size ratio following Equation \ref{EQ2} with $\theta=15\degree$ and introduced by shortening the interaction length scale between large and small disks. Importantly, the simulated Brownian dynamics depend only on interactions between disks and with the equilibrated bath, while neglecting any effects of electron beam perturbation. The observed agreement therefore indicates that the observed particle mobility does not rely on peculiarities of the electron beam interaction, but instead depends on common structural features of the experimental and simulated monolayers.

As a measure of particle mobility, in Figure~\ref{F4}c we plot the root-mean-square (RMS) displacement $\sqrt{\langle\delta r^2_S(\tau)\rangle}$, following evolution over a delay time $\tau$, over the full experimental range of size ratios. The delay time, which is further discussed in the Supporting Information, scales as $\tau \propto r_S r_L^2$ and reflects the Brownian time scale for small-particle diffusion through the large-particle lattice environment. We also plot the fraction of small particles, $f_{S, \rm hop}$ that hop between lattice sites within this time window, such that $|\delta r_S(\tau)|>2r_L$. In both experiment and simulation, small-particle mobility (blue) shows a crossover from a mobile, percolating phase at low size ratios to an immobile, trapped phase at higher size ratios. The mobile phase is characterized by frequent site-to-site hopping of the small-particle impurities leading to a divergent RMS displacement at late times. In the trapped phase, the confinement of small particles to individual sites suppresses long-range transport. For comparison, we note that the large-particle RMS displacement remains small for all size ratios and does not exceed the lattice spacing.

The enhancement in small-particle mobility can be understood by considering the non-additive configuration of the monolayer. Small particles are able to hop between hollow sites only when they are small enough to fit between large-particle gaps at bridge sites. Because they lie in separate planes, sufficiently small particles are able to pass through the channel formed above the large-particle contact point. As depicted in the inset of Figure~\ref{F4}c, this criterion is satisfied when $r_{LS}<r_L/2$. Through substitution from Equation \ref{EQ1}, we obtain:
\begin{equation}
(1+\sigma)(1+\Delta)<1 \label{EQ3}.
\end{equation}
Because of the well-defined functional relationship between $\sigma$ and $\Delta$ in Equation \ref{EQ2} we can fully parameterize this inequality in terms of $\sigma$, obtaining a numerical threshold of $\sigma\sim0.24$ for this system, as indicated by the dashed vertical line in Figure~\ref{F4}c, in close agreement with the observed crossover point.

Following this model, the predicted crossover is a direct consequence of the non-additive geometry. Notably, there is no solution to the inequality in Equation \ref{EQ3} in the case of $\Delta=0$, suggesting that the observed mobile phase is inaccessible to strictly planar systems. Previous investigations have found that particle motion becomes increasingly hindered with increasing surface coverage, approaching kinetic arrest in the dense limit.\cite{hunter_physics_2012, thorneywork_self-diffusion_2017,cui_transition_2017} To better understand the effects of non-additivity, we repeated the colloidal simulations using \emph{additive} WCA particles across the full range of size ratios. As shown in the inset of Figure~\ref{F4}d, the additive system corresponds to a monolayer geometry with $\theta=90\degree$ where all particles are attached to the interface at the same plane. In the plots, no crossover to a mobile phase is seen in measurements of either the RMS displacement or $f_{S, \rm hop}$ in the additive simulations. We still observe a marginal increase in the small-particle RMS displacement at low size ratios due to the expanded free area for small particles to explore within each hollow site, but displacements plateau at late times and do not exceed the lattice spacing. In the additive geometry, small particles are no longer able to slide over the gaps between large particles, and contact points therefore obstruct hopping pathways between hollow sites, restricting small-particle mobility for all size ratios.

\begin{figure*}
\includegraphics[width=17.8cm]{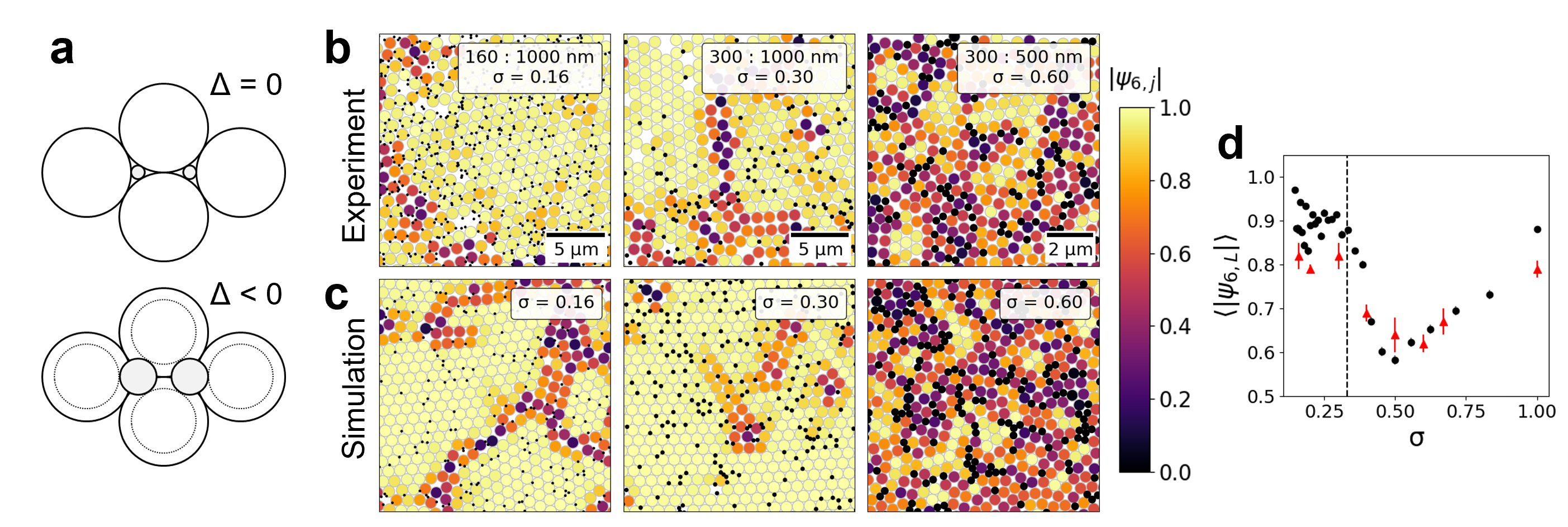}
\caption{\label{fig5} Lattice structure measurements. (a) Close-packed configuration of interstitial impurities of additive ($\Delta=0$) and non-additive packing geometries ($\Delta<0$). (b) Bidisperse monolayer images over a series of three size ratios $\sigma=0.16$, $0.30$, and $0.60$ with the large particles false colored by the magnitude of the hexagonal bond order parameter $|\Psi_{6,j}|$ and small particles colored black. (c) Snapshots of simulated non-additive WCA particles at the same size ratios as in (b). (d) Plot of the average large-particle hexagonal order $\langle|\Psi_{6,L}|\rangle$ as a function of size ratio measured from experiment (red triangles) and simulation (black circles).
}
\label{F5}
\end{figure*}

\textbf{\emph{Packing Constraints for Large-Particle Ordering.}} The structure of the underlying large-particle lattice environment defines the landscape for small-particle dynamics. In the absence of small particles, a monodisperse collection of nanoparticles at the fluid interface assembles to form a hexagonal lattice that maximizes surface density, as shown in Figure S3. During monolayer formation, the concurrent nucleation of multiple crystalline domains results in a polycrystalline structure with competing grain orientations. For mixtures including small particles, the question of optimal 2D packing becomes significantly more complex, and a wide variety of close-packed configurations exists depending on the composition and size ratio of the mixture. In practice, we do not obtain true close-packed structures as the propensity for maximizing surface coverage is mitigated by thermal fluctuations, interparticle interactions, and kinetic trapping, among other factors. Still, close-packing remains a useful heuristic, and to describe our observations, we pay special attention to the packing geometry where a single small particle fits exactly within the interstices of hexagonally-packed large particles, as depicted in Figure~\ref{F5}a. We focus on this case as it accommodates the small particle impurities while preserving the hexagonal symmetry of the large-particle sublattice. For additive systems, this packing occurs at a so-called ``magic'' size ratio of $\sigma=\frac{2}{\sqrt{3}}-1\simeq 0.15$ that corresponds to the case where $2 r_{LS}$ is equal to the distance between the center and vertex points of an equilateral triangle with side length $2 r_L$.\cite{likos_complex_1993} In non-additive systems, similar packing rules apply, but particle overlap allows larger particles to fit within the interstitial space, introducing a correction factor to $r_{LS}$, resulting in a magic ratio that we derived as
\begin{equation}
(1+\sigma)(1+\Delta)=\frac{2}{\sqrt{3}} \label{EQ4}.
\end{equation}
Substituting Equation \ref{EQ2}, into this expression and using $\theta = 15\degree$ we obtain a numerical threshold of $\sigma \sim 0.33$ for non-additive close-packing. Thus negative non-additivity shifts the stability of this close-packed configuration to higher values of $\sigma$.

In close-packed systems, when the size ratio deviates from this ``magic” value, shear deformation of the large-particle sublattice leads to a reduction of hexagonal symmetry.\cite{likos_complex_1993} We quantify the extent of deformation by measuring the large-particle hexagonal bond order parameter $\Psi_{6,j}$, defined for each large particle $j$ as:
\begin{equation}
\Psi_6,_j=\frac{1}{N_{\rm nn}}\sum_{k=1}^{N_{\rm nn}} e^{6i\theta_{jk}},   \label{EQ5}
\end{equation}
where $N_{\rm nn}$ is the number of nearest neighbor particles and $\theta_{jk}$ is the angle of the bond vector linking to the $k$-th neighbor. For this calculation, nearest neighbors are uniquely defined through Delaunay triangulation of the large-particle coordinates with a maximum center-to-center separation of 3 $r_L$. The order parameter is complex valued, such that the magnitude $|\Psi_{6,j}|$, ranging from 0 to 1, indicates the degree of local hexagonal order, while the phase corresponds to the grain orientation. Figure~\ref{F5}b compares large-particle order in experimental monolayers over a series of three size ratios. In each panel, large particles are colored according to their individual $|\Psi_{6,j}|$; small particles are not included for this analysis and have been colored black. Snapshots taken from equilibrated simulations for these size ratios display similar structural organization, as shown in Figure~\ref{F5}c. In the $\sigma=0.16$ and 0.30 examples, which fall below the threshold size ratio for close packing, large particles form an ordered polycrystalline monolayer, with ordered domains separated by disordered grain boundaries. The organization of the lattice in the $\sigma=0.30$ case, just below the threshold, does not differ significantly from that of $\sigma=0.16$. In both cases, the structure of the large-particle sublattice is not disrupted by the inclusion of small-particles and resembles monodisperse assembly. The lattice accommodates small-particles within vacant interstitial hollow sites while preserving the underlying hexagonal symmetry. By contrast, in the $\sigma=0.60$ monolayer we find that the ordering of large particles is significantly disrupted. In this regime, deformation is unavoidable, and hexagonal packing is frustrated by the presence of small particles. The monolayer maintains direct contacts between neighboring particles resulting in a generally amorphous random close-packed structure. Although the system exhibits no long-range order, patchy crystallite grains with local hexagonal symmetry are distributed throughout the monolayer, corresponding to regions that exclude small particles.\cite{ebert_partial_2009}

To assess structural properties across the full range of size ratios, in Figure~\ref{F5}d we average over all large particles to obtain the mean magnitude $\langle|\Psi_{6,L}|\rangle$ for each monolayer. Experimental uncertainties indicate the image-to-image standard deviation for different fields of view of the same ionic liquid droplet surface. In both experiment and simulation we observe a sharp decrease in hexagonal order crossing over the threshold size ratio of $\sigma\sim0.33$. At $\sigma<0.33$ where small particles are able to occupy the interstitial space,  $\langle|\Psi_{6,L}|\rangle$ remains relatively constant at a value comparable to monodisperse packing. This phase resembles a quasi-2D interstitial solid solution with partial occupation of lattice sites.\cite{van_der_meer_diffusion_2017} Although the hexagonal order achieved in simulation at $\sigma<0.33$ is systematically higher than in experiment due to the formation of larger polycrystalline grains, the similarity of ordering to monodisperse packing in each case confirms the absence of strain due to small-particle impurities. For $\sigma>0.33$, $\langle|\Psi_{6,L}|\rangle$ sharply decreases reaching a minimum at  $\sigma\sim0.50$ and then gradually increases approaching monodispersity. Small-particles no longer fit within the interstitial hollow of three large particles, and deformation of the large-particle lattice is unavoidable. At size ratios approaching $\sigma=1.0$, small particles can act as substitutional impurities in the hexagonal lattice, similarly relaxing lattice frustration for mixtures with low size asymmetry. 

\begin{figure*}
\includegraphics[width=12cm]{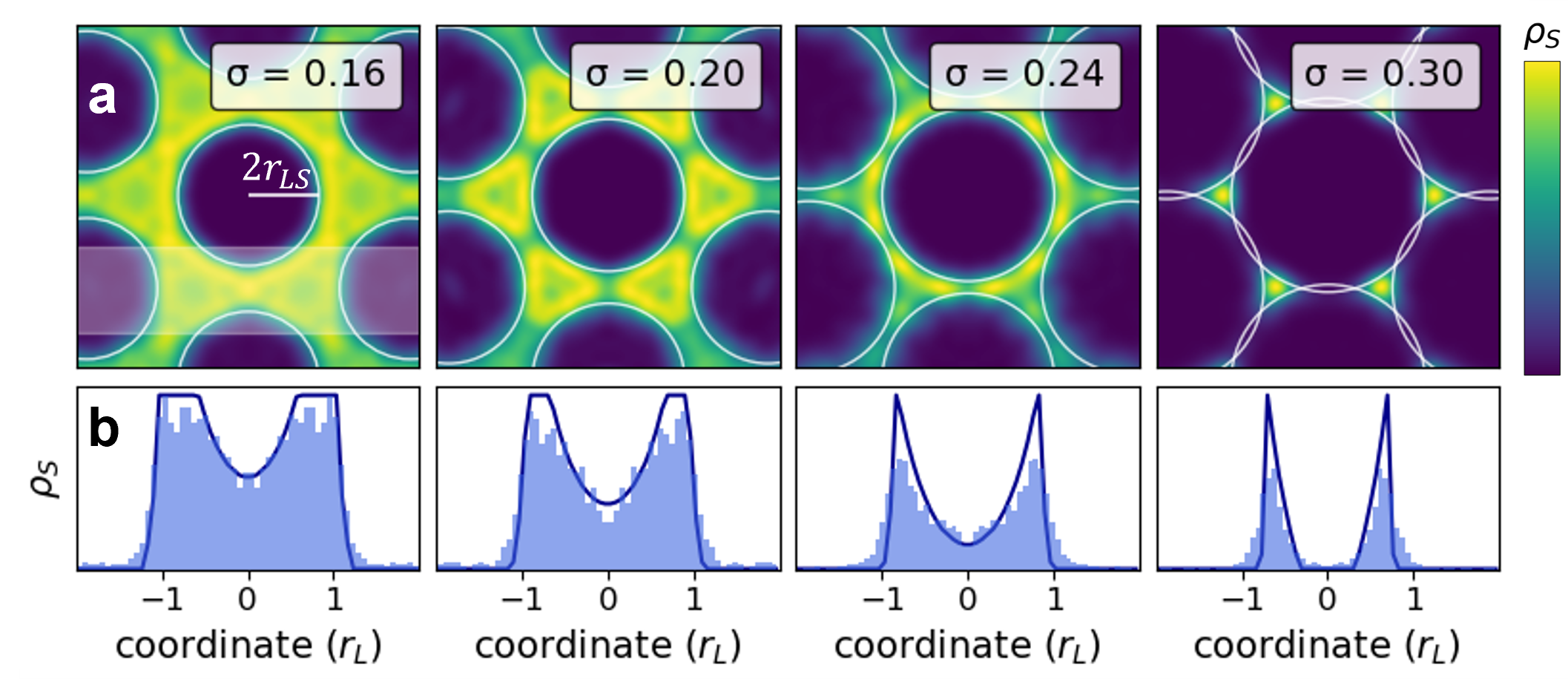}
\caption{\label{fig6} Small-particle distribution maps. (a) Real-space maps showing the spatial distribution of small particles averaged over ordered large-particle lattice sites for a series of a size ratios  $\sigma=0.30$, $0.24$, $0.20$, and $0.16$. The color-scale indicates the local probability density of small particles, with yellow being high density and blue being low density. Small-particles occupy large-particle lattice interstices and are found to be excluded from circles with a radius $2 r_{LS}$ centered on large particle centers, with the borders of the excluded area indicated by white circles. (b) Density maps projected along the hopping coordinate between two lattice sites, as indicated by the annotated region in the first panel of (a). These plots compare the measured small particle density (pale blue) from (a) to the total interstitial free area (dark blue) as determined from the area of the excluded lattice sites.
}
\label{F6}
\end{figure*}

\textbf{\emph{Static Distributions Predict Small-Particle Mobility.}} In monolayers with size ratios below $\sigma\sim0.33$, small particle impurities are distributed throughout the interstices of the ordered large-particle sublattice. In this regime, small particles do not contribute to the mechanical stability of the monolayer, as they are not large enough to be in contact with each of their immediate large-particle neighbors. As observed in our mobility measurements, they are therefore free to move within the interstitial space, occupying a range of positions in the empty region defined by the large-particle lattice\cite{koeze_mapping_2016}. At size ratios below $\sigma\sim0.24$ the extent of small-particle motion increases further due the opening of continuous transport pathways for percolation. Real-space maps of the small-particle probability density $\rho_S$ are generated by plotting the experimentally measured small-particle positions relative to a central large particle, rotated to obtain a consistent unit cell orientation. Small-particle positions are then mapped over each large-particle lattice hollow site to account for the six-fold rotational symmetry of the lattice. By accumulating small particles over hundreds of unit cells, we obtain a representative sampling of their local distribution, and the resulting $\rho_S$ maps are plotted in Figure~\ref{F6}a. The full details of this analysis are discussed in the Supporting Information and presented in Figure S4. Because distributions are measured using the first SEM image obtained on each region, we do not expect SEM charging or other imaging artifacts to have influenced the particle configuration.

Figure~\ref{F6}a shows that at each size ratio, $\rho_S$ is distributed within the interstitial space within volume-excluded large particle lattice sites. White circles indicate the borders of the excluded area inaccessible to small-particle centers, extending a distance $2 r_{LS}$ from the large particle center in the contact plane. In the $\sigma=0.16$ monolayer, hollow sites are joined by continuous open pathways enabling free transport of small particles.  At this size ratio, the significant height offset between large and small particles effectively lowers the overall particle density in the contact plane. With increasing size ratio, $r_{LS}$ increases, expanding the excluded area and shrinking the pathways for small-particle percolation through lattice. In the $\sigma=0.30$  monolayer, the excluded area of adjacent large particles overlaps, causing $\rho_S$ to be separated into disconnected hollow sites. Altogether, the small-particle distributions obtained from static images recapitulate the dynamics observed in SEM movies. The size ratio where the excluded area no longer overlaps and transport pathways first appear coincides with the $\sigma\sim0.24$ threshold determined from observations of particle dynamics. This strong correspondence between our measurements of the initial structure and the ensuing dynamics further demonstrates the capability for sensitive SEM imaging of liquid samples under minimally perturbative conditions.

To quantify transport probabilities, in Figure~\ref{F6}b we compute $\rho_S$ projected along the site-to-site hopping coordinate from the corresponding data in Figure~\ref{F6}a for each size ratio, \emph{i.e.}, hops from the center of one large-particle hollow site to the next. The small-particle density distribution (pale blue) is plotted alongside the free area due to large-particle volume exclusion (dark blue), each integrated over the cross-section of the sampled region annotated by the white band in the $\sigma=0.16$ panel of Figure~\ref{F6}a. Comparison of these curves allows us to evaluate the relative contributions of geometric and non-geometric effects on the local monolayer structure. While the probability density generally follows from these geometric predictions, differences arise due to interparticle interactions and lattice dynamics. Along the hopping coordinate $x$, $\rho_S$ is minimized at the narrowest point between large particles and is maximized at hollow sites. We therefore calculate a hopping barrier $E_{\rm a}$ for transport between sites following
\begin{equation}
E_{\rm a}=-k_{\rm B}T \ln\frac{\min(\rho_{S}(x))}{\max(\rho_{S}(x))}, \label{EQ6}
\end{equation}

\noindent where $k_{\rm B}$ is the Boltzmann constant and $T$ is temperature. Measured activation energies increase with increasing size ratio and are included in Table 1. Notably, in the $\sigma=0.30$ case, we observe non-zero probability for small particles along the full hopping coordinate, leading to a barrier of $3.4\pm0.3$ $k_{\rm B}T$. The corresponding static geometric model predicts that large particles should fully obstruct the hopping pathway, resulting in an infinite barrier. In practice, this geometric constraint is relaxed due to out-of-plane fluctuations, heterogeneity between lattice sites, and the elastic response of the large-particle lattice. Additionally, dynamics of the host lattice have previously been shown to facilitate impurity transport in interstitial colloids.\cite{tauber_anomalous_2016} Similarly, for each monolayer, we observe that distribution fringes extend a finite distance into the excluded region, effectively lowering the hopping barrier. At smaller size ratios where transport pathways are wider, this effect becomes less significant, and experimentally measured barriers converge with geometric predictions.

\begin{table}
  \caption{Activation energies for small-particle site-to-site hopping.}
  \label{tbl:T1}
  \begin{tabular}{ll}
    \hline
    $\sigma$  &  $E_{\rm a}/k_{\rm B}T$  \\
    \hline
    0.16  &  $0.60\pm0.02$   \\
    0.20  &  $1.19\pm0.08$  \\
    0.24  &  $1.45\pm0.07$  \\
    0.30  &  $3.42\pm0.29$ \\
    \hline
  \end{tabular}
\end{table}

In a hard-disk system, where particles interact only through area exclusion, we would expect small particles to be evenly distributed over the region of accessible microstates. Instead, $\rho_S$ shows local variability within each hollow site, with enhanced probability for small particles close to neighboring large particles. This pattern, which is most apparent in the $\sigma=0.20$ distribution, is indicative of short-range attractive interactions between large and small particles. We expect that these forces primarily arise from the capillary attraction of interfacially-bound particles, as electrostatic interactions are effectively screened by the ionic liquid solvent.\cite{kim_assessing_2019, bischak_charging-driven_2020} While the simulations reported here rely on purely repulsive interactions, we find that the observed phase behavior is robust to the inclusion of modest attractive interactions with well depths of up to roughly 5 $k_{\rm B}T$. Simulations with stronger interparticle attraction result in irreversible aggregation and hindered dynamics. Comparison of simulation results for different interparticle potentials is included in Figure S5.

\section*{Discussion}

Taken together, our dynamic and structural observations allow us to classify the bidisperse monolayers into three distinct phase regimes. Phase boundaries for these regimes are determined from the thresholds obtained from Equations \ref{EQ3} and \ref{EQ4}.
For $\sigma<0.33$, the monolayer forms a large-particle polycrystalline lattice, where small particles are disconnected from the large-particle packing network and fluctuate within the interstitial space. Within this regime, for $\sigma<0.24$, small particles are able to percolate freely through the empty space defined by the large-particle lattice. For $0.24<\sigma<0.33$, however, the accessible space becomes partitioned into disconnected tricuspid cages, causing the range of small-particle motion to be localized within individual hollow sites. In monolayers with  $0.33<\sigma<0.67$, small-particles no longer fit within hollow sites and therefore establish mechanical contact with the packing network. In this range, large and small particles together form a randomly packed assembly with negligible particle mobility. Previous investigation of bidisperse monolayers with $\sigma=0.41-0.78$ demonstrated the formation of a jammed state with no detectable particle rearrangements over several hour periods.\cite{gao_bidisperse_2020} Mixtures in this range of size ratios are commonly employed as glass formers, characterized by amorphous structure and slow dynamic time scales.\cite{konig_experimental_2005, koeze_mapping_2016} Although our observations are also consistent with particle jamming, we note that experimental monolayer densities are lower than typical critical packing fractions of $\phi_T=0.81-0.89$ for random close packing in 2D.\cite{zaccone_explicit_2022} Altogether, this behavior is consistent with geometric predictions for the interaction of non-additive hard disks.

\begin{figure*}
\includegraphics[width=12cm]{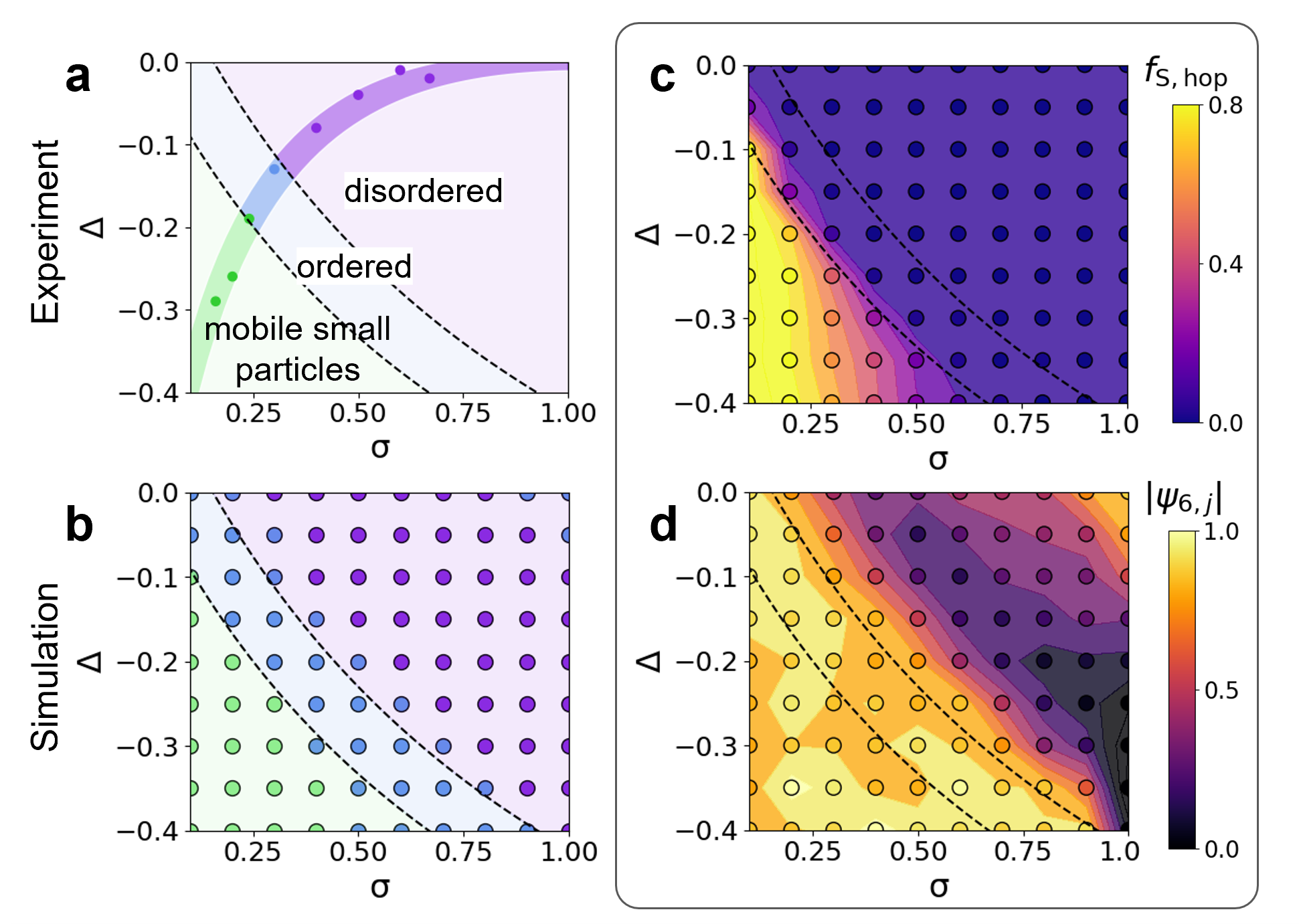}
\caption{\label{fig7} Overview of observed phase behavior. (a) Plot of the predicted phase behavior of non-additive monolayers with varying size ratio $\sigma$ and non-additivity $\Delta$. A small-particle mobility crossover is predicted at $(\sigma+1)(\Delta+1)=1$ and a structural crossover is predicted at $(\sigma+1)(\Delta+1)=2/\sqrt{3}$, as indicated by the dashed lines. Experimental data follows a single slice through phase space. (b) Plot of phase behavior measured from simulations of WCA particles with varying $\sigma$ and $\Delta$ and $\chi_S=0.33$. Observed phases generally correspond with predicted crossovers. (c) Contour plots showing the small-particle hopping fraction $f_{S,{\rm hop}}$ (upper panel) and average large-particle hexagonal order $\langle|\Psi_{6,L}|\rangle$ (lower panel) measured from each simulation. Threshold values of $f_{S,{\rm hop}}>0.40$ for small-particle mobility and $\langle|\Psi_{6,L}|\rangle>0.75$ for large-particle ordering were used to determine the simulated phase behavior.
}
\label{F7}
\end{figure*}

Over the series of small/large particle ratios studied, non-additivity plays an essential role in determining the observed phase behavior. In particular, the small-particle mobility crossover relies on the formation of a continuous interstitial matrix for small-particle percolation. Such interconnectivity is only possible when small-particles are able to slide through the void space between large particles and is therefore inaccessible to additive systems. While enabled by the underlying large-particle lattice structure, the enhanced small-particle mobility also influences the organization of the large particles. Because differently sized particles interact in separate contact planes, non-additivity effectively decouples their dynamics, allowing reconfiguration of the large-particle sublattice with order and packing density equivalent to analogous monodisperse systems. In additive monolayers, contacts between large and small particles mutually suppress their mobility leading to kinetic arrest far from equilibrium.\cite{cui_transition_2017}

While non-additivity is here determined by the interfacial geometry, it can also occur in binary systems due to “soft” interactions from ligand intercalation, solvent-mediated interactions, or nanocrystal faceting, among other effects.\cite{silvera_batista_nonadditivity_2015} Under most circumstances, however, there is no straightforward correspondence between the non-additivity and size ratio. In this geometric case, the functional relationship between $\sigma$ and $\Delta$ allows us to fully parameterize the monolayer’s phase behavior, but also makes it challenging to disentangle their independent contributions. As shown in Figure~\ref{F7}a, the experimental measurements follow a path through phase space, crossing thresholds defined by Equations \ref{EQ3} and \ref{EQ4}. We can test the observed behavior more generally by performing simulations of non-additive WCA particles with constant $\chi_S=0.33$ and systematically varying $\sigma$ and $\Delta$ over a larger range of the phase space. Along $\sigma= 1$, non-additivity is introduced by arbitrarily dividing the monodisperse disks into two subpopulations at separate effective heights. This condition has been previously realized experimentally, by sandwiching colloidal particles between parallel plates with variable separation.\cite{pieranski_two-dimensional_1980, han_geometric_2008} The resulting simulated phase diagram is shown in Figure~\ref{F7}b, in which equilibrated monolayers have been classified into disordered (purple), ordered (blue), or mobile (green) phases according to threshold values of $f_{S,{\rm hop}}$ and $\langle|\Psi_{6,L}|\rangle$. This phase diagram, measured from simulations over the full range of $\sigma$ and $\delta$, qualitatively agrees with both our experimental findings and our geometric predictions, capturing the three expected phase regimes of the system. Differences between the predicted phase boundaries, shown by the dashed curves, and the simulated data may be understood by considering  the corresponding contour maps of small-particle mobility, $f_{S,{\rm hop}}$, and large-particle order, $\langle|\Psi_{6,L}|\rangle$, shown in Figures 7(c) and (d), respectively. In Figure 7(c), small-particle hopping occurs more frequently than predicted at low $\sigma$ because lattice dynamics create transient pathways that accommodate hops for smaller particles.  In Figure 7(d), lattice ordering with $\langle|\Psi_{6,L}|\rangle>0.8$ is consistently observed at values of $\sigma$ slightly higher than predicted by the order/disorder boundary (upper dashed curve). Lattice elasticity, structural heterogeneity, and dynamic reorganization of the lattice together contribute to relax packing constraint and allow the lattice to retain hexagonal symmetry at higher size ratios than predicted. Despite not accounting for these effects, the simple geometric model successfully captures much of the observed phase behavior. Furthermore, several observations are found that are not included in this simplified representation: In regions near $\sigma=1$ and $\Delta=0$ at the top right corner of Figure 7(d), a second ordered phase is formed where impurities act as substitutional defects without disrupting the lattice. Additionally, monodisperse simulations in the range  $\Delta=-0.20$ to $-0.40$ in the lower right of Figure 7(d), have low hexagonal symmetry, but include regions containing ordered square lattices. This square-symmetric phase assembles to maximize the interlayer packing efficiency of confined particles at intermediate layer thicknesses.\cite{pieranski_two-dimensional_1980, curk_layering_2012}

Although the experiments have examined only one slice of the accessible phase space for non-additive binary monolayers, they highlight the importance of non-additivity for understanding the assembled phase behavior. By modifying the surface chemistry or roughness of the particles in order to change $\theta$, the relationship between $\sigma$ and $\Delta$ could be tuned to fully explore the effects of non-additivity. These modifications may be readily achieved following strategies used to control the wettability of colloidal particles for stabilizing emulsions.\cite{xiao_tailoring_2018} In particular, monolayers with arbitrary $\sigma$ and $\Delta$ could be prepared through independent control over the contact angle of each component. Looking forward, the unique attributes of non-additive monolayers achieved through interfacial confinement suggests a practical avenue for engineering exotic monolayer morphologies, including Kagome lattices\cite{chen_directed_2011, salgado-blanco_non-additive_2015}, binary superlattices,\cite{zhou_discovery_2022} or quasicrystals.\cite{fayen_self-assembly_2023} As these geometric assemblies do not rely on specific chemical interactions, the nanoparticle building blocks could be further functionalized to introduce desired monolayer properties. Interfacial particle assemblies have found broad utility including as surfactants for stabilizing emulsions,\cite{shi_nanoparticle_2018} catalysts for biofuel reactors,\cite{crossley_solid_2010} and templates for the fabrication of nanostructured films.\cite{lotito_approaches_2017} The packing rules studied here for quasi-2D morphologies may also be extended to the organization of fully 3D colloidal crystals through layer-by-layer assembly.\cite{velikov_layer-by-layer_2002}

\section*{Conclusion}

Via minimally-invasive electron microscopy, we have investigated the influence of non-additivity on the phase behavior of interfacially confined bidisperse colloidal monolayers. SEM imaging of liquid droplets without encapsulation enables direct access to the ionic liquid/vacuum interface for complete tracking of particle coordinates in space and time. The interfacial attachment geometry, with large and small particle equators bound at separate vertical planes, can be naturally represented as a 2D system of non-additive hard-disks, in which non-additivity is determined by the mixture’s size ratio. Non-additivity is critical for understanding the observed monolayer properties, which can be classified into disordered, ordered, and mobile phase regimes. In particular, at low size ratios where non-additivity is most significant, the monolayer exhibits small-particle transport through site-to-site hopping that would not be possible in a strictly planar, additive geometry. The structural and dynamic properties are interdependent, as the enhanced mobility of the non-additive system facilitates equilibration of the lattice. Altogether, the observed behavior is predicted from particle packing constraints of the non-additive geometry and is recapitulated through molecular dynamics simulations of non-additive disks.

Looking beyond binary size mixtures of silica particles, multicomponent or heterogeneous colloidal systems may be tuned to achieve diverse and sophisticated functionality. For example, in semiconducting particle monolayers, size and spectral heterogeneity lead to exciton funneling for light-harvesting.\cite{nguyen_imaging_2017} By virtue of using electron microscopy, one could obtain the electron beam-induced optical emission (cathodoluminescence) to demonstrate resonant energy transfer within such monolayers, collecting light emitted from some particles due to electron beam excitation of other, spectrally distinct, neighboring ones, all during monolayer evolution at an ionic liquid interface. Here, we have identified non-additivity in colloidal mixtures as a critical ingredient for describing their assembly and phase dynamics. Although non-additivity here is directly determined by the binary size ratio, it is a general feature of many multicomponent systems and may arise due to ligand intercalation, particle faceting, and ionic screening, among other factors. Irrespective of its origin, control over non-additive interactions offers a powerful and complementary strategy for tuning the properties of colloidal assemblies.

\section*{Methods}
\subsection*{Colloidal Monolayer Preparation.}
Monodisperse colloidal solutions were prepared by concentrating and redispersing aqueous stock solutions of silica nanospheres with bare silanol surface chemistry (160-1000 nm diameter, 10 mg $\rm mL^{-1}$, nanocomposix) in 1-ethyl-3-methylimidazolium ethyl sulfate ($\rm EMIM^+:EtSO_4^-$) ionic liquid. The monodisperse solutions had a corresponding size dispersity of 4.7 $\%$ (160 nm), 8.0 $\%$ (200 nm), 4.7 $\%$ (300 nm), 3.8 $\%$ (500 nm) and 2.2 $\%$ (1000 nm). These solutions were placed in a vacuum chamber for 1 hour to remove excess water. Each mixture was then prepared with a concentration of 50 mg $\rm mL^{-1}$ and a 2:1 small:large number density. An additional 25 vol.$\%$ glycerol was then added to each solution. We have empirically found that the addition of glycerol improves structural ordering of the resulting colloidal monolayers, potentially by reducing electrostatic screening with respect to neat IL to prevent colloidal aggregation during assembly. Next, 3 $\rm \mu L$ droplets were deposited onto cleaned $\sim 1 \: \rm cm \times 1 \: \rm cm$ Si wafer substrates (Virginia Semiconductor). Substrates were cleaned by solvent rinses with isopropyl alcohol, acetone, and distilled water followed by 2 min $\rm O_2$ plasma cleaning. Prior to imaging, droplets were stored under an inert $\rm {N_2}$ environment for at least 12 h to allow for monolayer equilibration.

\subsection*{Scanning Electron Microscopy.}
Droplets on Si wafers were grounded with a copper clip and loaded into the chamber of a Zeiss Gemini SUPRA 55 S2 SEM. Imaging was performed using an accelerating voltage of 3 keV and a beam current of $15\pm2$ pA. Each SEM movie and image was acquired at a new sample region to mitigate the effects of beam exposure. The beam dose, which depended on the size of the imaging field of view, was varied over a range of $5-16 \: e^- \, \rm nm^{-2} \, \rm s^{-1}$.

\subsection*{Single Particle Tracking and Analysis.}

SEM data were analyzed with custom python code. Individual particles were identified from image data using a Laplacian of Gaussians filter, with particle sizes determined from the standard deviation of each feature.  From measurement of stationary colloidal lattices, we estimate a center positional uncertainty of 25 nm. Following automated size classification, binary particle assignments were manually confirmed for each image. Features from each frame were linked into time-dependent trajectories using the trackpy package, which implements the Crocker-Grier algorithm.\cite{crocker_methods_1996} Trajectories were drift corrected using the large-particle sublattice as a stable reference.

\subsection*{Molecular Dynamics Simulations.}
Molecular dynamics simulations of binary particles with variable size ratio and non-additivity were conducted using LAMMPS.\cite{thompson_lammps_2022} Each simulation was performed in two dimensions with periodic boundary conditions. Particle interactions were described using a WCA potential to produce a steep short-range repulsion, and non-additivity was incorporated by shortening the WCA interaction length scale for large-small interactions. Simulations were carried out in Lennard-Jones units, with the fundamental time step determined from the diffusive time scale of the large particles. To ensure consistent particle density across the range of size ratios, simulations were performed in the isothermal-isobaric ensemble in a flexible simulation box, using a Langevin thermostat to maintain constant temperature and a Nosé-Hoover barostat to maintain constant pressure. Constant pressure conditions reflect our experiments where the imaging field-of-view is surrounded by a dense network of particles. For each size ratio, the initial conditions were determined for a system of 900 particles with a number fraction of $\chi_S = 0.33$ by steadily increasing the system pressure to its final equilibrium value of $P = 0.1$ in reduced Lennard-Jones units. Following initialization, we confirm equilibration by verifying constant system properties and dynamics independent of the sampling time.

\section*{Conflicts of Interest}
There are no conflicts to declare.

\section*{Acknowledgments}

We thank A. Das for valuable discussions and input on simulations and A. Omar for helpful comments on the manuscript. We thank E. Wong, E. S. Barnard, D. F. Ogletree, and S. Aloni at the Molecular Foundry for assistance with SEM. This work has been supported by STROBE, A National Science Foundation Science \& Technology Center under Grant No. DMR 1548924. The SEM imaging at the Lawrence Berkeley Lab Molecular Foundry was performed as part of the Molecular Foundry user program, supported by the Office of Science, Office of Basic Energy Sciences, of the U.S. Department of Energy under Contract No. DE-AC02-05CH11231. N.S.G. acknowledges an Alfred P. Sloan Research Fellowship, a David and Lucile Packard Foundation Fellowship for Science and Engineering, and a Camille and Henry Dreyfus Teacher-Scholar Award.

\begin{suppinfo}

Discussion of the effects of beam perturbation; analysis of the dependence of the SEM accelerating voltage on particle imaging; derivation of the relevant dynamic timescale for small-particle transport; analysis of the packing of monodisperse assemblies; procedure for generating small-particle distribution maps; and discussion of the effects of interparticle attraction on molecular dynamics simulations.

\end{suppinfo}


\bibliography{maintext_bib.bib}

\end{document}